\documentclass[a4paper]{jpconf}
\usepackage{graphicx}
\usepackage{lscape,longtable}
\usepackage{multirow}
\usepackage{epsfig}
\usepackage{graphicx}
\usepackage{epstopdf}
\usepackage{afterpage}
\begin{document}
\title{Future prospects of di-jet production constraining $\Delta g(x)$ at low $\mathbf{x}$ at STAR at RHIC}
\author{Bernd Surrow (for the STAR Collaboration)}
\address{Temple University, Department of Physics, 1900 N. 13th Street, Philadelphia 19122, USA}
\ead{surrow@temple.edu}
\begin{abstract}
One of the main objectives of the high-energy spin physics program at RHIC at BNL is the precise determination of
the polarized gluon distribution function, $\Delta g(x)$. 
Polarized $\vec{p}+\vec{p}$ collisions at $\sqrt{s}=200\,$GeV and at $\sqrt{s}=500\,$GeV at RHIC provide an unique way to probe the 
proton spin 
structure. Inclusive measurements, such as inclusive 
jet and hadron production, have so far been the prime focus of various 
results at $\sqrt{s}=200\,$GeV constraining $\Delta g(x)$
for $0.05<x<0.2$. 
A recent global analysis provides
for the first time evidence
of a non-zero value of the gluon polarization 
$\int_{\tiny 0.05}^{\tiny 1}\Delta g (x)\, dx \,(Q^{2}=10\,{\rm GeV}^{2}) = 0.20^{+0.06}_{-0.07}$.
First results of di-jet production at $\sqrt{s}=200\,$GeV
by the STAR collaboration will allow a better constraint of the underlying event kinematics.
Extending the current program to smaller values of $x$ is a key goal 
for the future high-energy spin physics program at RHIC. Forward di-jet production measurements at 
STAR beyond the current acceptance from $-1<\eta<+2$ to $+2.5<\eta<+4$, in 
particular those carried out at $\sqrt{s}=500\,$GeV, provides access to low $x$ values at the level of
$10^{-3}$ where current uncertainties of $\Delta g(x)$ remain very large. 
Those measurements will eventually be complemented by a future Electron-Ion
Collider facility probing $\Delta g(x)$ in polarized $\vec{e}+\vec{p}$ collisions.
\end{abstract}

\section{Introduction}

\vspace*{0.25cm}

\begin{figure}[tbh]
\centerline{\includegraphics[width=100mm]{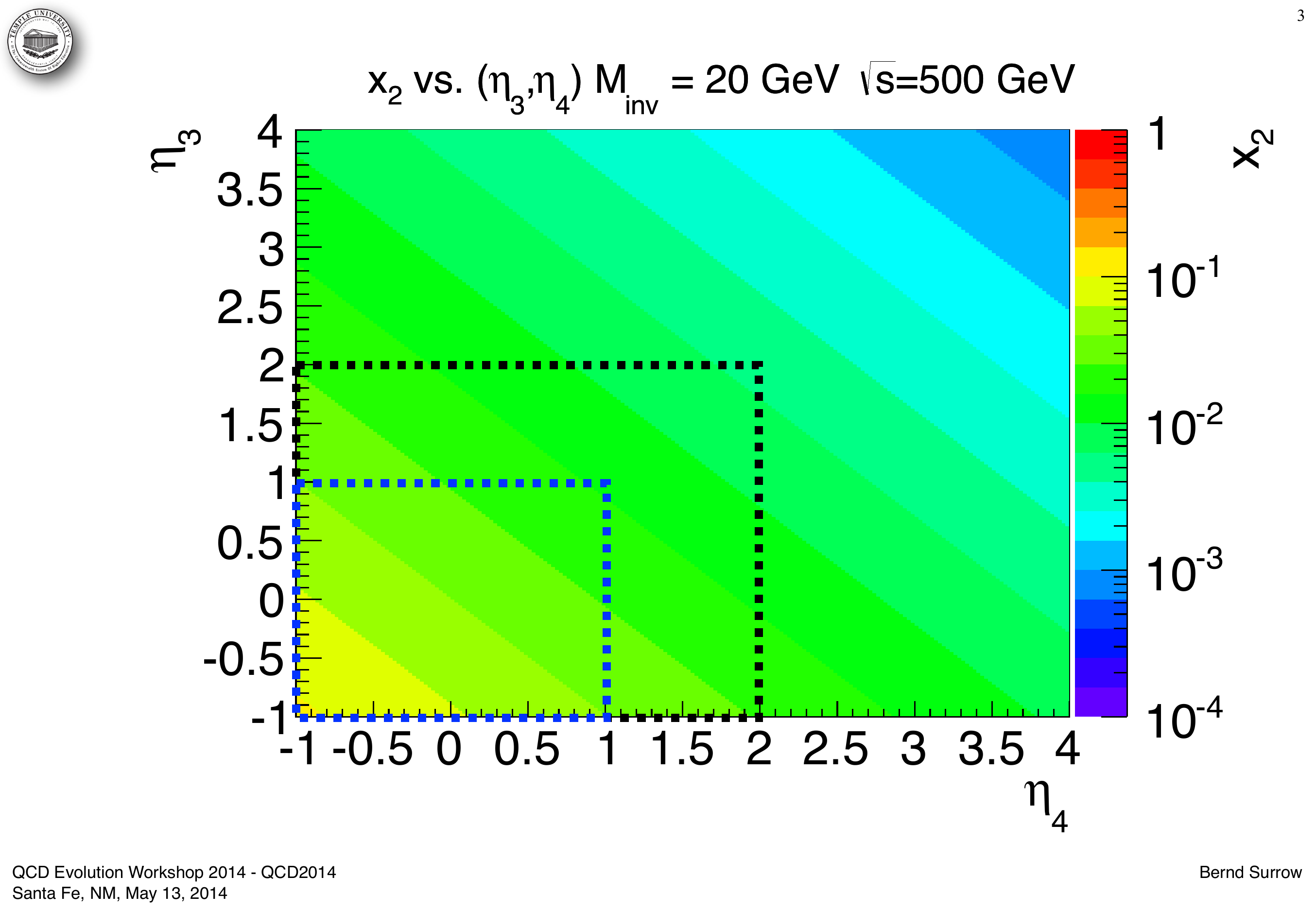}}
\caption{\label{fig:Kinematics}{\it Low-$x$ coverage ($x_{2}$) displayed as a color shade for di-jet final states 
showing $\eta_{3}$ versus $\eta_{4}$. The black and blue dashed lines indicate the region for which results have been released (black) and the region for which STAR has been fully instrumented (blue) without yet releasing any results as of now.}}
\end{figure}

High energy polarized $\vec{p}+\vec{p}$ collisions at a center-of-mass energy of $\sqrt{s}=200\,$GeV and at $\sqrt{s}=500\,$GeV at RHIC at BNL provide an unique way to probe the proton spin structure using very well established processes in high-energy physics, both experimentally and theoretically. The measurement of the gluon polarization at RHIC has been formulated as a key milestone of the US Nuclear Physics program \cite{US-NP}. A recent global analysis provides for the first time evidence of a non-zero value of the gluon polarization restricted to the current experimentally accessible kinematic range of $0.05 < x < 0.2$ \cite{ref1}. The gluon polarization program by the STAR experiment at RHIC is based on the measurement of various probes, such as inclusive hadron and jet production and di-jet production in polarized proton-proton collisions. The full exploitation of the STAR Barrel and Endcap Electromagnetic Calorimeter (BEMC / EEMC) is crucial for this analysis effort. The STAR collaboration has published measurements of the longitudinal double-spin asymmetry $A_{LL}$ for inclusive jet production \cite{ref2,ref3,ref4} and neutral pion production \cite{ref5,ref6}. The most significant result of inclusive jet production at 
$\sqrt{s}=200\,$GeV is based on data collected during the 2009 running period (Run 9) \cite{ref7}. This result is dominating the impact on the gluon polarization extracted in a recent updated global analysis \cite{ref1}. The first significant di-jet measurement is also based on data collected during Run 9 \cite{ref8}. Both, the asymmetry measurement for inclusive jet production and di-jet production are clearly pointing towards a larger gluon polarization compared to the first global analysis which included results form the Run 2005/2006 RHIC polarized $\vec{p}+\vec{p}$ running periods of inclusive jet and hadron production. While inclusive jet measurements provide a strong sensitivity on the gluon polarization, the actual measurement effectively integrates over a wide range in $x$. Future measurements require additional means to provide a direct sensitivity to the actual $x$ dependence of $\Delta g (x)$. 

\begin{figure}[tbh]
\centerline{\includegraphics[width=120mm]{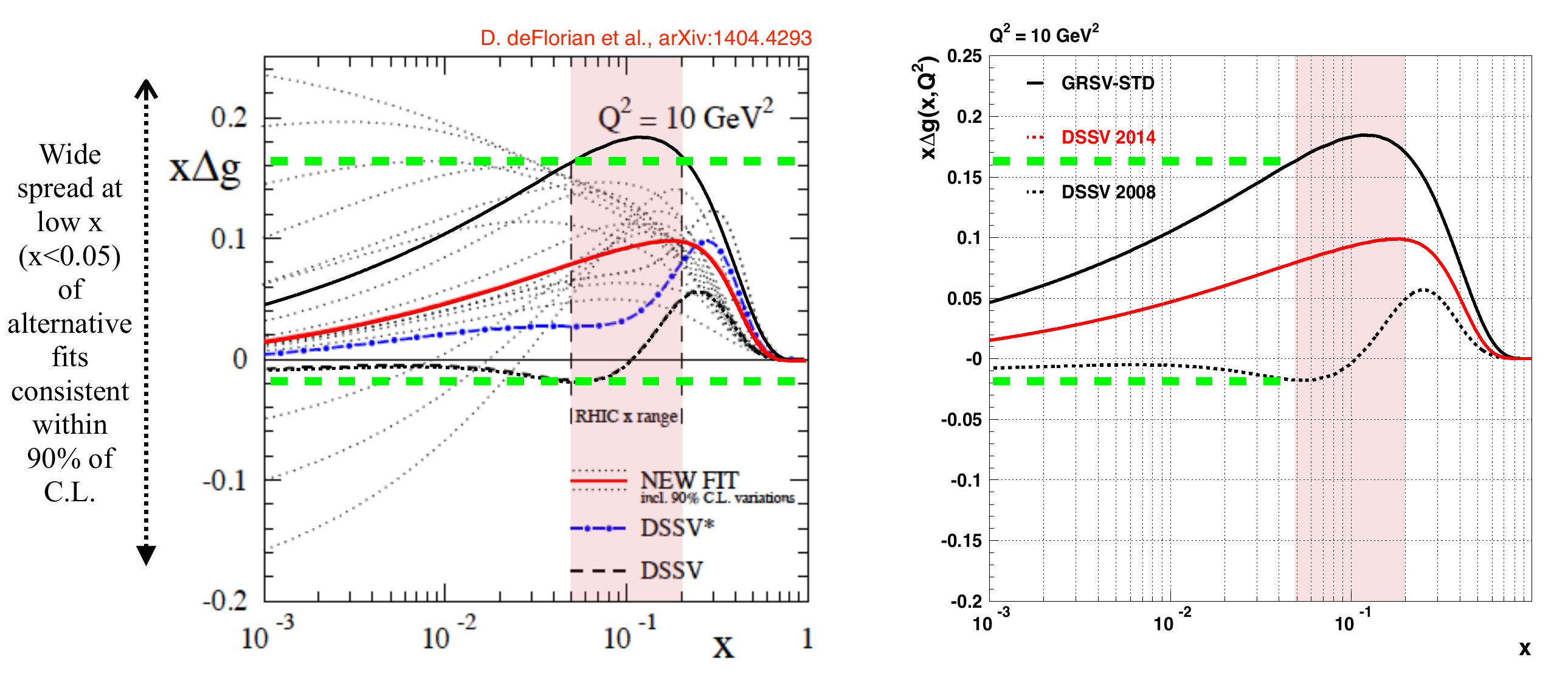}}
\caption{\label{fig:pdf}{\it Comparison of DSSV (2008) \cite{ref10} and GRSV-STD \cite{ref11} to DSSV (2014) \cite{ref1}.}}
\end{figure}

\section{Probing the gluon polarization at low x through di-jet production}

\vspace*{0.25cm}

\subsection{Kinematics}

\vspace*{0.25cm}

Di-jet production at STAR will allow for a better constraint of the underlying event kinematics to constrain the shape of  
$\Delta g (x)$. The invariant di-jet mass M is related to the product of the initial partonic $x$ values, i.e. 
$x_{1}\cdot x_{2}$, whereas the pseudo-rapidity sum $\eta_{3}+\eta_{4}$ is related to the ratio of the partonic $x$ values, i.e. 
$x_{1} / x_{2}$, since $M^{2}=s\cdot x_{1}x_{2}$ and $\eta_{3}+\eta_{4}=\ln(x_{1}/x_{2})$ based on elementary four-vector kinematics.  Measurements at both $\sqrt{s}=200\,$GeV and at $\sqrt{s}=500\,$GeV are preferred to maximize the kinematic region in $x$. The wide acceptance of the STAR experiment permits reconstruction of di-jet events with different topological configurations, i.e. different $\eta_{3} / \eta_{4}$ combinations, ranging from symmetric ($x_{1} \simeq x_{2}$) partonic collisions to 
asymmetric ($x_{1} \ll x_{2}$ or $x_{1} \gg x_{2}$) partonic collisions. It is in particular the access to the large 
$\eta_{3} / \eta_{4}$ region which allows one to probe gluons in QCD processes at very small $x$-values. The current acceptance region of the STAR experiment includes the EAST ($-1.0 < \eta < 0.0$) and WEST ($0.0 < \eta < 1.0$) coverage of the BEMC along with the EEMC ($1.09 < \eta < 2.0$). The proposed new forward region would cover a nominal range in $\eta$ of $2.5 < \eta < 4.0$. Figure~\ref{fig:Kinematics} illustrates the lower $x$-range covered for a di-jet final state of $M=20\,$GeV displaying the actual $x$ value as a color shade as a function of $\eta_{3}$ and $\eta_{4}$ for the current 
($-1.0 < \eta < 2.0$) and future ($2.5 < \eta < 4.0$), proposed forward acceptance region . The black and blue dashed lines indicate the region for which results have been released (black) and the region for which STAR has been fully instrumented (blue) without yet releasing any results as of now. One can clearly see that the current $\eta$ range allows one to probe a region in x of approximately $0.05<x<0.2$. Extending the current region to include the EEMC region of $1.09 < \eta < 2.0$ would extend the $x$ range to at least $10^{-2}$. Additional instrumentation involving a Forward Calorimeter System (FCS) at forward rapidity for $2.5 < \eta < 4.0$ would allow an extension to $x$ values as low as $10^{-3}$ \cite{LOI-STAR-Forward}.

\subsection{Projected performance}

\vspace*{0.25cm}

The kinematic range in $x$ along with the projected uncertainties for di-jet longitudinal double-spin asymmetries $A_{LL}$ have been evaluated using a MC framework at NLO level \cite{ref9}. Those studies are based on the following assumptions:
Jet efficiencies for the BEMC EAST and WEST regions and EEMC region are constrained by data ranging from $~30\%$ at $
p_{T}=10\,$GeV and rising to $~90\%$ ($95\%$) at $p_{T}=20\,$GeV ($p_{T}=30\,$GeV). For the forward acceptance region, it is assumed that a hadronic calorimeter system is available with expected larger efficiencies at the level of $90\%$.
Asymmetric di-jet cuts in $p_{T}$ have been chosen to be $5\,$GeV and $8\,$GeV.
The largest systematic uncertainty is expected to be dominated by the relative luminosity ($R$) measurement relevant for the actual asymmetry measurement.  An uncertainty of $\delta R = 5\cdot 10^{-4}$ has been assumed which is in fact the quoted value for the Run 9 inclusive jet analysis \cite{ref8}. This value is therefore considered to be a conservative estimate for the purpose of the di-jet $A_{LL}$ projections presented here.
A beam polarization of $60\%$ and a total delivered luminosity of $1000\,$pb$^{-1}$ have been assumed with a ratio of $2/3$ for the ratio of recorded to delivered luminosity. The assumed delivered luminosity would require that a full RHIC running period is devoted to $500\,$GeV operation. 

\begin{figure}[t]
\centerline{\includegraphics[width=120mm]{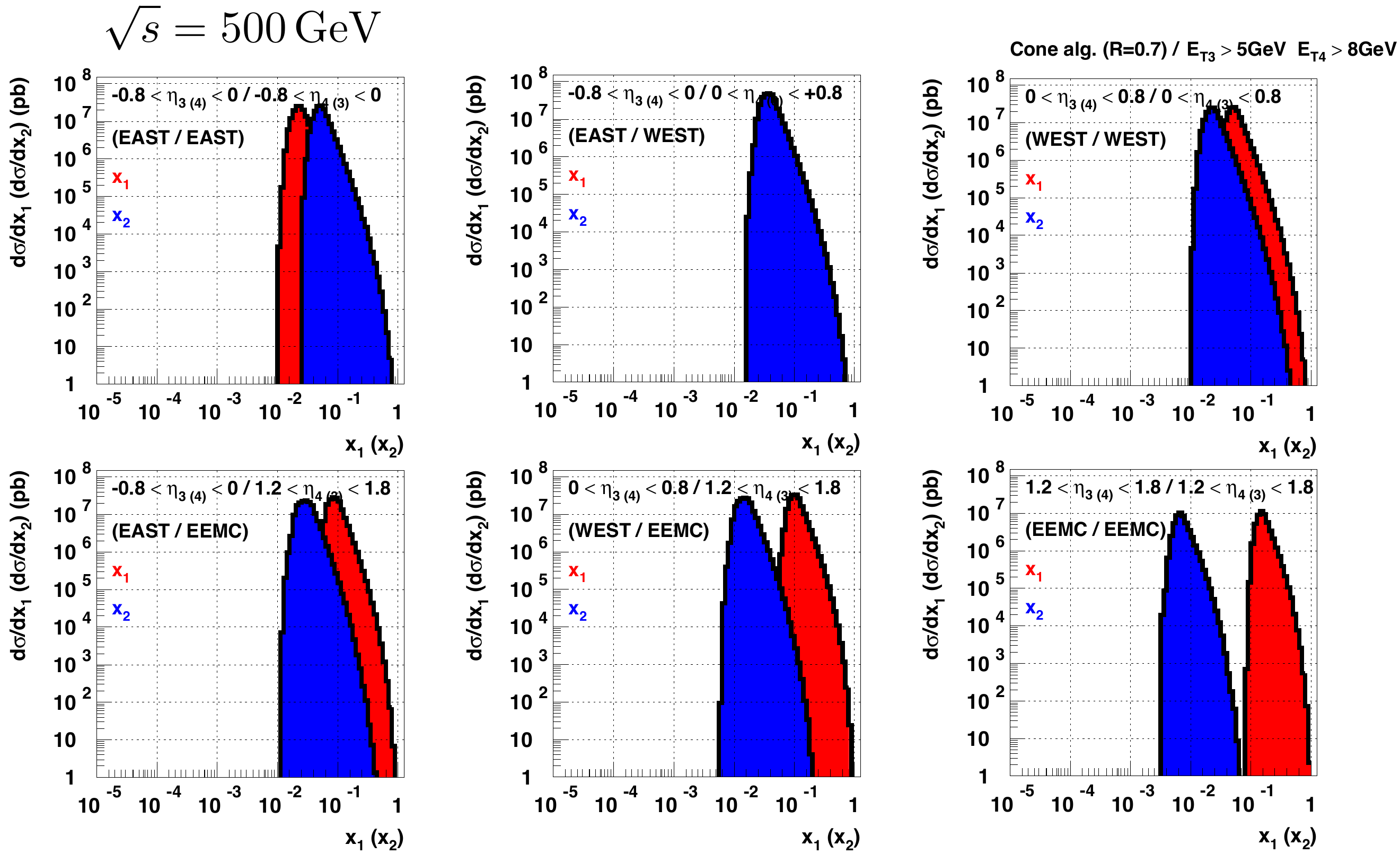}}
\caption{\label{fig:x1x2-central-500}{\it $x_{1}$ / $x_{2}$ range for the current STAR acceptance region 
in $\eta$ of $-1 < \eta < 2$.}}
\end{figure}

\begin{figure}[b]
\centerline{\includegraphics[width=120mm]{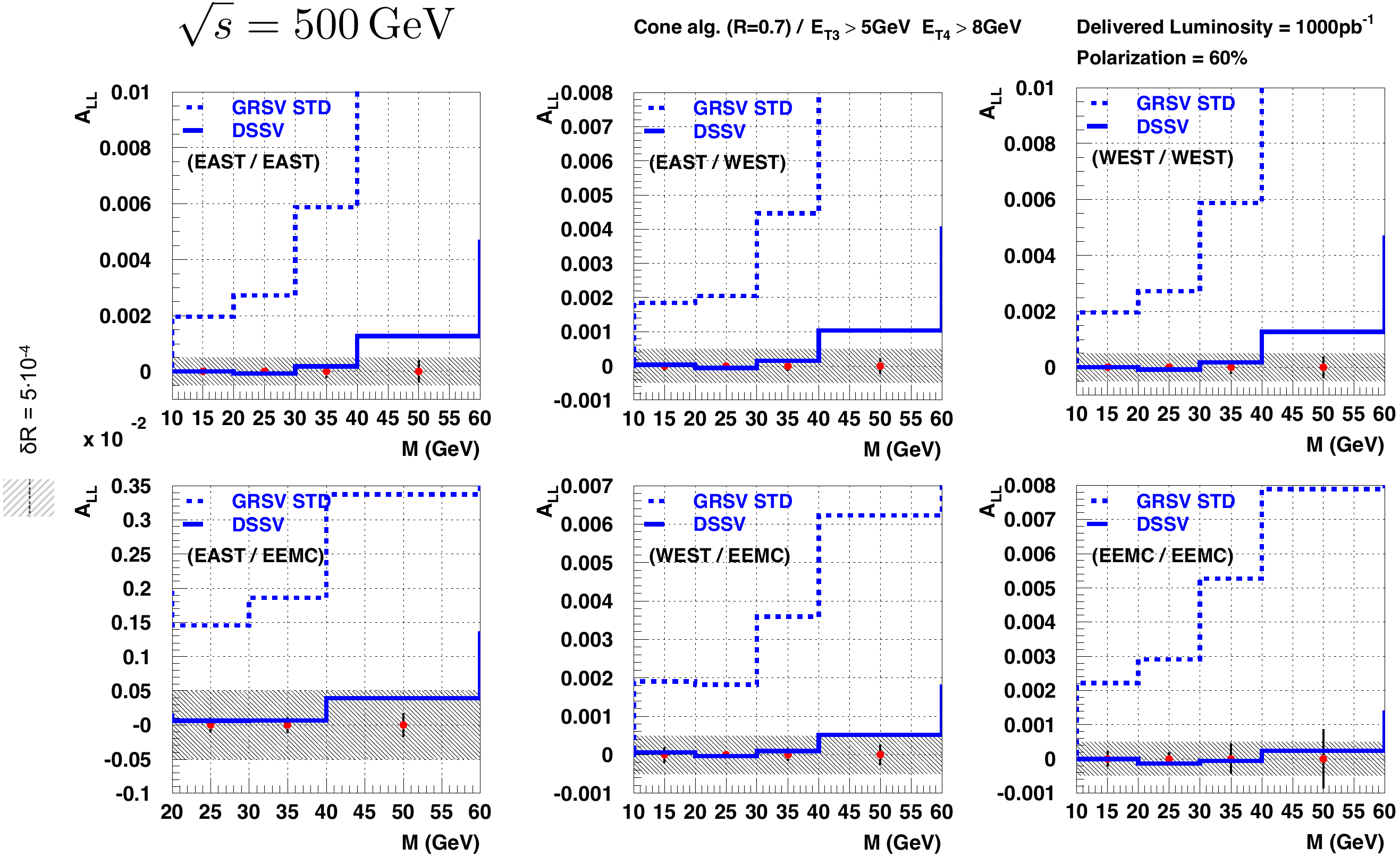}}
\caption{\label{fig:Central-ALL-500}{\it $A_{LL}$ NLO calculation for $-1 < \eta < 2$ together with projected uncertainties. }}
\end{figure}

Figure~\ref{fig:pdf} shows a comparison of DSSV (2008) \cite{ref10} and GRSV-STD \cite{ref11} in comparison to DSSV (2014) \cite{ref1}. Both DSSV (2008) and GRSV-STD have been used for all projection studies of $A_{LL}$. The recent DSSV (2014) analysis released an uncertainty envelope based on fits at $90\%$ C.L. which are shown as dotted lines in Figure~\ref{fig:pdf}. The range for DSSV (2008) and GRSV-STD is smaller than the 
$90\%$ C.L. envelope for $x<0.05$. The range in $A_{LL}$ for DSSV (2008) and GRSV-STD is therefore not covering the full uncertainty range. Figure~\ref{fig:x1x2-central-500} shows the $x_{1} / x_{2}$ range for the current STAR acceptance region in $\eta$ of $-1 < \eta < 2$ for six different topological di-jet configurations. The EEMC / EEMC configuration would permit probing $x$ values around $10^{-2}$ and thus well below the current x region for which experimental results have been achieved. 
Figure~\ref{fig:Central-ALL-500} shows the actual asymmetries $A_{LL}$ as a function of the invariant mass $M$ for the same topological di-jet configurations. The theory curves at NLO level have been evaluated for DSSV (2008) and GRSV-STD  where DSSV (2008) refers to the first global fit result including polarized $\vec{p}+\vec{p}$ collision data from RHIC. 
Both the statistical and also the systematic uncertainties are much smaller compared to the separation between DSSV and GRSV-STD. 

\begin{figure}[t]
\centerline{\includegraphics[width=120mm]{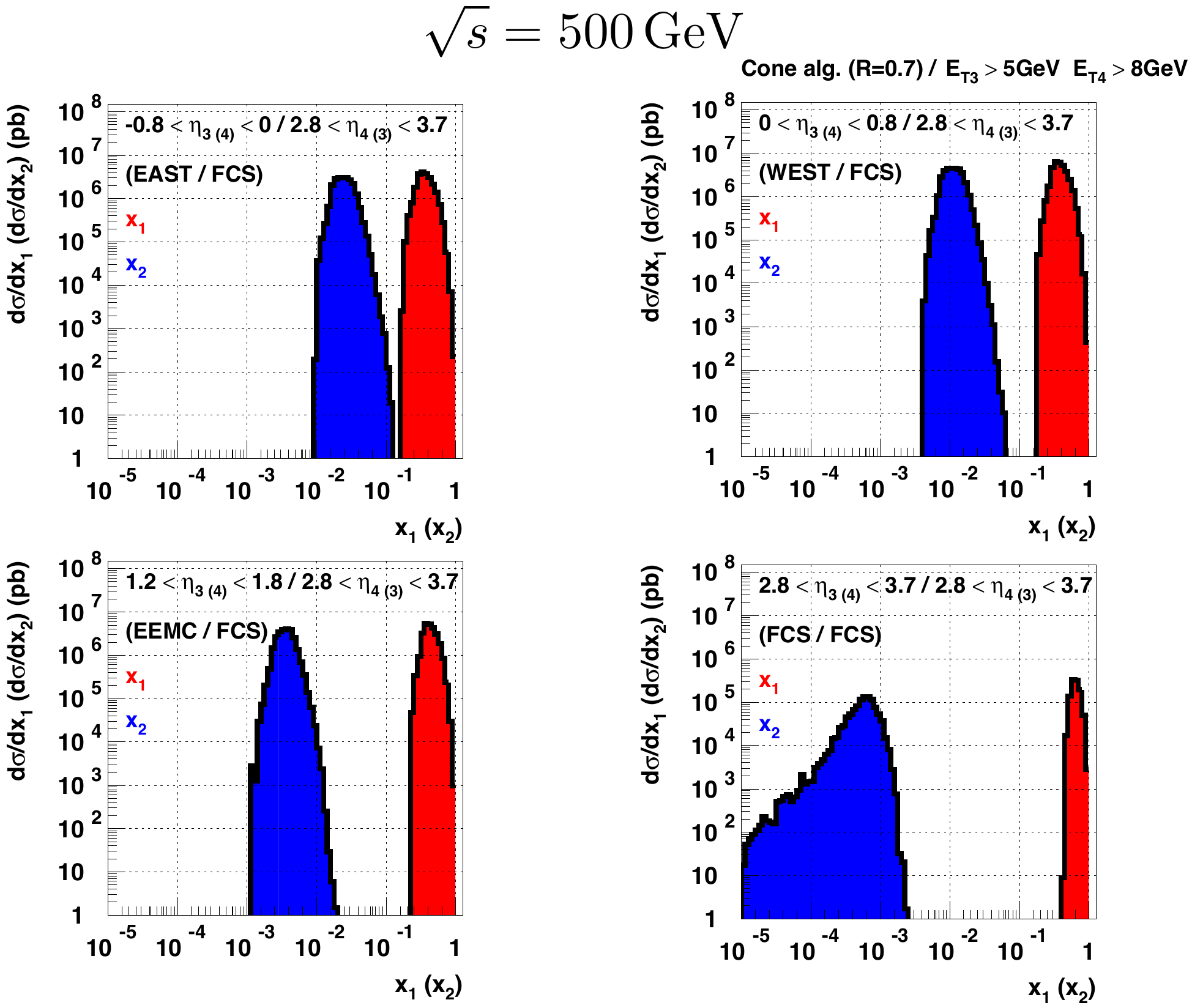}}
\caption{\label{fig:x1x2-forward-500}{\it $x_{1}$ / $x_{2}$ range for the forward STAR acceptance region in $\eta$ 
of $2.8 < \eta < 3.7$.}}
\end{figure}

\begin{figure}[b]
\centerline{\includegraphics[width=120mm]{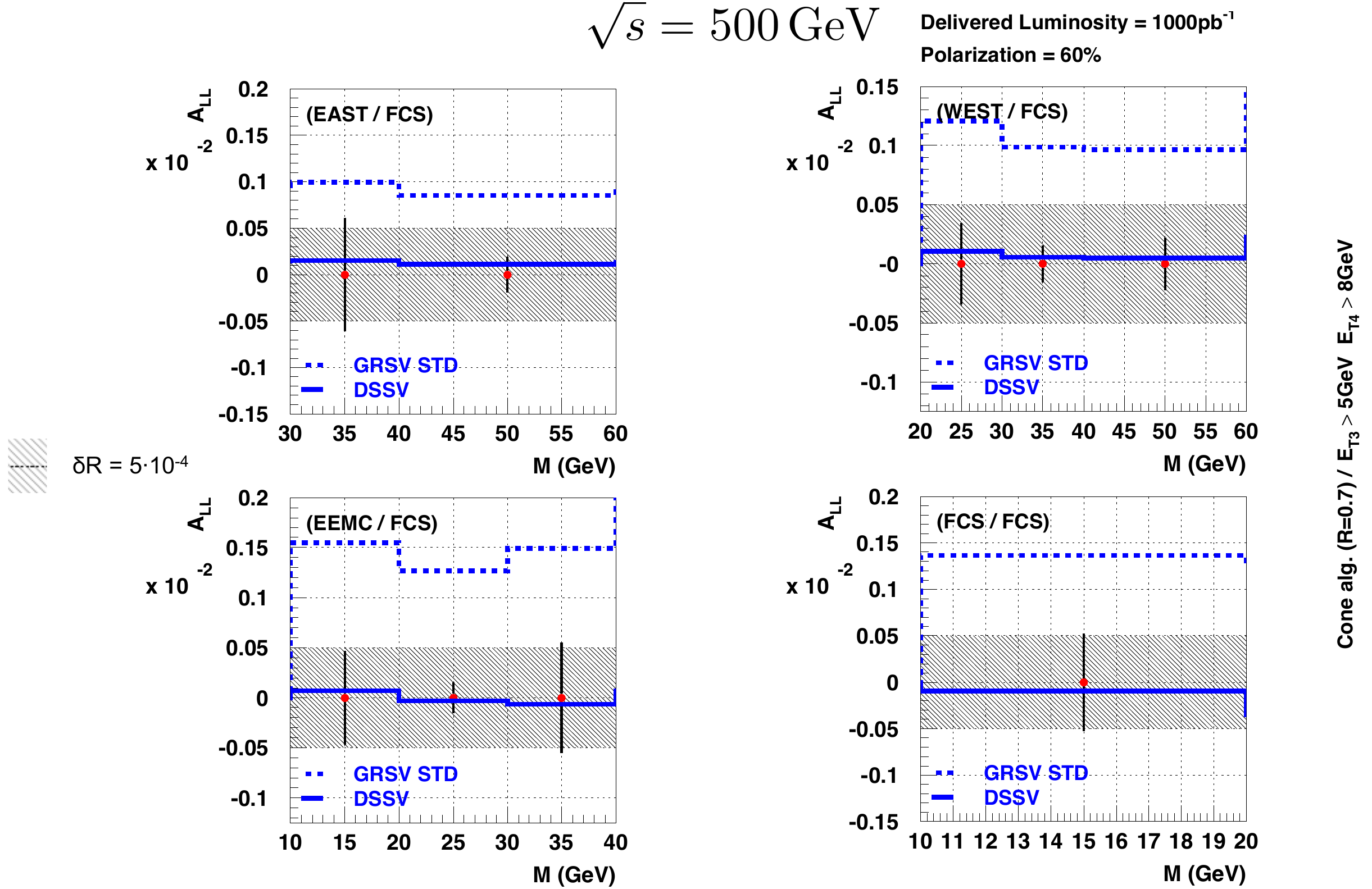}}
\caption{\label{fig:Forward-ALL-500}{\it $A_{LL}$ NLO calculations for 
$2.8 < \eta < 3.7$ together with projected uncertainties.}}
\end{figure}

Figure~\ref{fig:x1x2-forward-500} shows the x coverage for four topological di-jet configurations involving at least the forward system labeled as FCS in combination with either EAST, WEST, EEMC and FCS. It is in particular the EEMC / FCS and FCS / FCS configurations which would one allow to probe $x$ values as low as $10^{-3}$. Figure~\ref{fig:Forward-ALL-500} shows the actual asymmetries $A_{LL}$ as a function of the invariant mass M for the same topological di-jet configurations. The theory curves at NLO level have been evaluated for DSSV \cite{ref10} and GRSV-STD \cite{ref11} where DSSV refers also to the first global fit result to include data from RHIC in polarized $\vec{p}+\vec{p}$ collisions. The systematic uncertainty, which is assumed to be driven by the relative luminosity uncertainty of $\delta R = 5\cdot 10^{-4}$, is clearly dominating over the size of the statistical uncertainties. Any future measurements in this topological configuration of very forward measurements would clearly benefit from improved relative luminosity measurements. This also includes the expected benefit from a fully commissioned spin flipper system at RHIC. However, even with the assumed Run 9 relative luminosity result, a measurement of strong impact can be achieved comparing the $A_{LL}$ curves for DSSV and GRSV-STD evaluated at NLO level. Such measurements would probe for the first time the unexplored kinematic region around $10^{-3}$ in $x$ prior to a future Electron-Ion Collider program. The recent global analysis released by the DSSV group concludes that low-$x$ data are `badly needed' \cite{ref3}. A measurement of forward di-jet production would provide such measurements. The size of the actual cross-section in the FCS / FCS configuration is substantially smaller compared to all other configurations. Accessing this configuration is crucial to allow access to values in $x$ below $10^{-3}$. 

This study assumes only a forward calorimeter system. The impact of a tracking system based on either silicon disks or GEM disks would have only a marginal effect on the $p_{T}$ reconstruction considering the STAR axial magnetic field configuration. However, a tracking system is expected to improve the actual localization and separation of jets, in particular for FCS / FCS configuration. Furthermore a forward tracking system should be part of a forward jet calibration strategy based on particle finding capabilities. High rate capability and efficiency are essential performance measures in particular for background rejection. More detailed simulations beyond basic fast simulations are required to address these questions. A forward hadronic calorimeter system is essential for efficient jet measurements. The first forward di-jet measurements at RHIC at $\sqrt{s}=500\,$GeV have been performed by the AnDY collaboration \cite{ref12}.

\section{Summary}

\vspace*{0.25cm}

In summary, forward di-jet production in combination with measurements of the current STAR acceptance region would allow to probe spin phenomena of gluons well below the currently accessible region of $0.05 < x < 0.2$. Such measurements would provide critical initial insight into the nature of the proton spin considering that gluons could contribute in a significant way to the spin of the proton. Such a program would pave a solid path towards a future Electron-Ion Collider and emphasize why such a facility is ultimately needed to probe the x dependence of the gluon polarization below $10^{-3}$ in $x$ with high 
precision \cite{ref13}.

\end{document}